\documentclass[showpacs,amsmath,nofootinbib,amssymb,twocolumn,superscriptaddress]{revtex4-1}

 \usepackage{graphicx}
\usepackage{array}
\usepackage{color}
\usepackage{dcolumn}
\usepackage{bm}
\usepackage{epsf}

\usepackage{amsmath} 

\newcommand{\beq}{\begin{equation}}
\newcommand{\eeq}{\end{equation}}
\newcommand{\bean}{\begin{eqnarray*}}
\newcommand{\eean}{\end{eqnarray*}\noindent}
\newcommand{\bea}{\begin{eqnarray}}
\newcommand{\eea}{\end{eqnarray}\noindent}

\begin{document}
\topmargin0in
\textheight 8.in 
\bibliographystyle{apsrev}

\title{Open issues in neutrino astrophysics}
\author{Cristina Volpe}
\email{volpe@apc.univ-paris7.fr}
\affiliation{AstroParticule et Cosmologie (APC), UMR 7164-CNRS, Universit\'e Paris Diderot-Paris 7, 10, rue Alice Domon et L\'eonie Duquet, 75205 Paris cedex 13, France}

\begin{abstract}
Neutrinos of astrophysical origin are messengers  produced in stars, in explosive phenomena like core-collapse supernovae, in the accretion disks  around black holes, or in the Earth's atmosphere.   Their fluxes and spectra encode information on the environments that produce them. Such fluxes are modified in characteristic ways when neutrinos traverse a medium. 
Here our current understanding of neutrino flavour conversion in media is summarized. The importance of this domain for astrophysical observations is emphasized. Examples are given of the fundamental properties that astrophysical neutrinos have uncovered, or might reveal in the future. 
\end{abstract}

\maketitle

\section{Introduction}

R. Davis' pioneering measurement of electron neutrinos emitted by the sun has opened the era of neutrino astronomy  \cite{Davis:1964hf}. 
A significant deficit, compared to the standard solar model predictions,
was rapidly measured, triggering several decades of experiments of solar neutrino experiments \cite{Hirata:1989zj,Abdurashitov:1999zd,Anselmann:1994cf,Arpesella:2008mt,Ahmad:2002jz} to investigate, if unknown neutrino properties, or solar model uncertainties \cite{Bahcall:1987jc,TurckChieze:1993dw,Adelberger:1998qm,Robertson:2012ib}, were at its origin. 
Electron antineutrinos produced by a massive star were first detected during the SN1987A explosion.  A burst was first seen in Kamiokande, the electron direction pointing to the Large Magellanic Cloud at 50 kpc from the Milky Way. The occurrence probability  of having 9 events per 10 seconds was determined to be less than 5.7 $ \times 10^{-8}$  \cite{Suzuki:2008zzf}. Altogether the Kamiokande, IMB, Baksan and Mont Blanc detectors observed about twenty events \cite{Hirata:1987hu,Bionta:1987qt,Alexeyev:1987,Aglietta:1987it}, in the 40 MeV energy range during 13 seconds (Figure 1).
Their angular dependence and energy distribution are reasonably consistent with expectations from core-collapse supernova simulations  (see e.g. \cite{Pagliaroli:2008ur} for a recent analysis). The detection of these events constitute the first experimental confirmation of  the predictions from supernova simulation.
In 2002, R. Davis and M. Koshiba were awarded the Nobel Prize, for "pioneering contributions to astrophysics, in particular for the detection of cosmic neutrinos". The prize was shared with R. Giacconi for "for pioneering contributions to astrophysics, which have led to the discovery of cosmic X-ray sources".
At the same epoch of the solar puzzle, experiments searching for proton decay were measuring an anomaly in the atmospheric neutrino background \cite{Hirata:1988uy,Aglietta:1988be,Fukuda:1994mc, Casper:1990ac, Berger:1990rd,Fogli:1996nn}. In 1998 the discovery of neutrino oscillations by the Super-Kamiokande experiment solved the atmospheric anomaly and brought a milestone in the solution of the "solar neutrino deficit" problem \cite{Fukuda:1998mi}. In fact, a deficit of upgoing  atmospheric muon neutrinos traversing the Earth was observed, compared to down-going  ones, that was shown to be consistent with the hypothesis that up-going muon neutrinos convert into tau neutrinos (Figure 2). This discovery has fundamental implications for high-energy physics, astrophysics and cosmology. For example, neutrinos from the sun 
give us direct confirmation of the energy production mechanisms in stars (see the seminal works \cite{Bethe:1939bt,gs}).

\begin{figure}
  \includegraphics[width=9.cm]{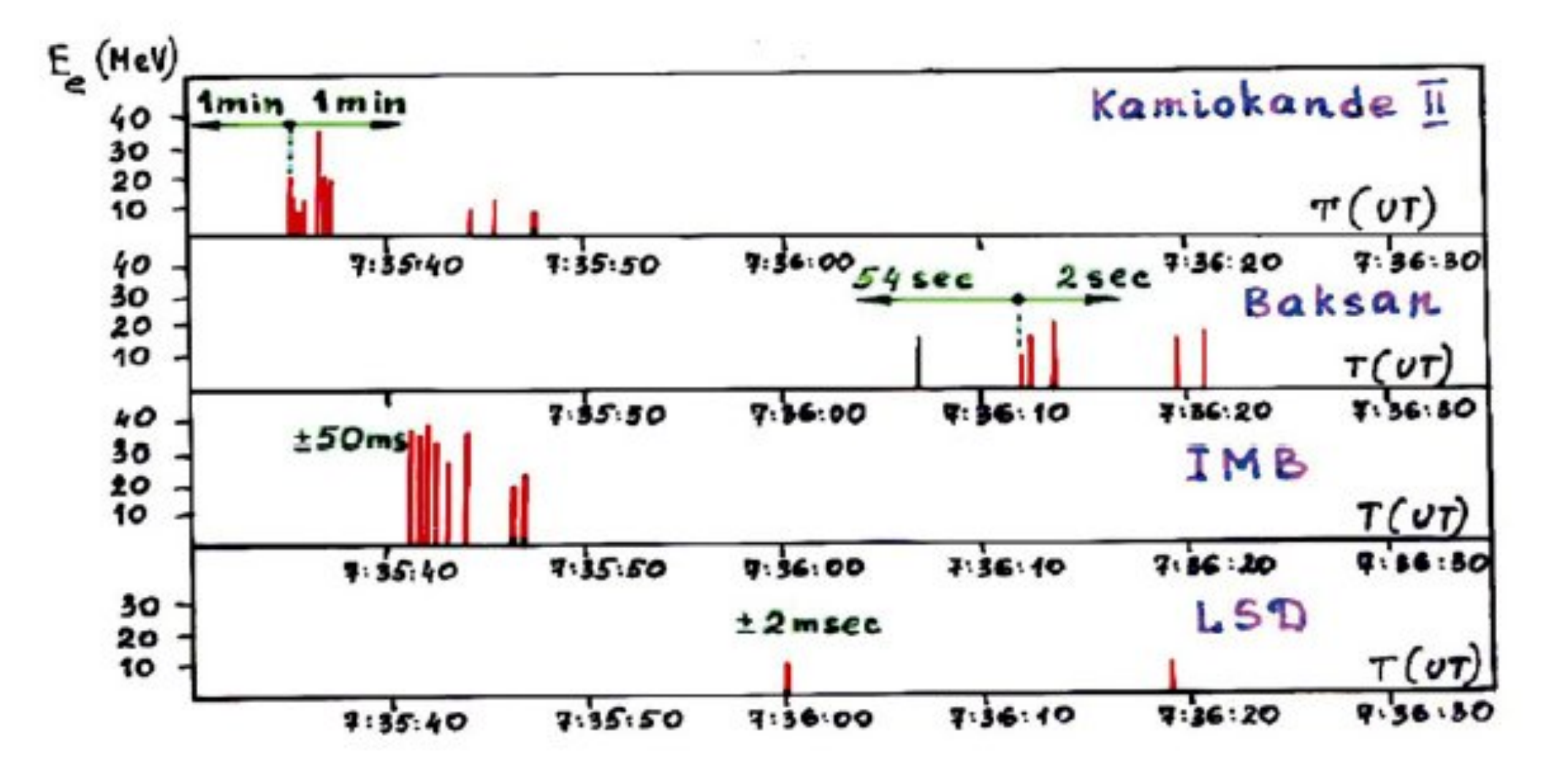}
  \caption{ Electron anti-neutrino events measured by the Kamiokande, Baksan, IMB and LSD experiments during the supernova 1987A, in the Large Magellanic Cloud \cite{Suzuki:2008zzf}.}
\end{figure}
The neutrino oscillation phenomenon occurs if neutrinos have non-zero masses and mixings. The possibility to have $\nu$-$\bar{\nu}$ oscillations was first pointed out by B. Pontecorvo in analogy with $K_0-\bar{K}_0$ mesons \cite{Pontecorvo:1957cp}. Neutrino oscillations require that the flavour $\nu_{\alpha}$ and the mass $ \nu_i$ basis are related by $\nu_{\alpha} = U^*_{\alpha i} \nu_i$ with $U$ being the Pontecorvo-Maki-Nakagawa-Sakata  (PMSN) unitary matrix \cite{Maki:1962mu}.  In the three-flavour framework the $U$ matrix reads
\begin{equation}\label{e:mnsp}
U=VD
\end{equation}
with a possible parametrization given by \cite{Beringer:1900zz}  :
\begin{equation}\label{e:V} 
V=
\left(\begin{array}{lll}
c_{12} c_{13} & s_{12} c_{13} & s_{13} e^{-i \delta}
 \\
 -s_{12} c_{23} - c_{12} s_{23} s_{13} e^{i \delta} &
 c_{12} c_{23} - s_{12} s_{23} s_{13} e^{ i \delta} &
 s_{23} c_{13}
 \\
 s_{12} s_{23} - c_{12} c_{23} s_{13} e^{i \delta} &
 -c_{12} s_{23} - s_{12} c_{23} s_{13} e^{i \delta} & c_{23} c_{13}
\end{array}\right)
\end{equation}
where $c_{ij}=\cos \theta_{ij}$ and $s_{ij}=\sin \theta_{ij}$ and
\begin{equation}\label{e:D}
 D=
\left(\begin{array}{ccc}
e^{-i\phi_1} & 0 & 0 \\
 0 & 1 & 0  \\
0 & 0  & e^{-i\phi_2}
\end{array}\right)
\end{equation}
The V matrix is analogous to the Cabibbo-Kobayashi-Maskawa matrix, in the quark sector, although the mixing angles are there measured to be small.  The $\phi_1, \phi_2$ are two extra phases that appear in case neutrinos are Majorana particles. Therefore the PMNS matrix depends on three mixing angles, two Majorana-type and one Dirac-type phases.  (Here we will not discuss the Majorana phases, since they do not influence neutrino oscillations in vacuum and in matter.) If the Dirac-type phase is non-zero, the PMNS matrix is complex, introducing a difference between the neutrino oscillation probability, and the corresponding  ones for antineutrinos, implying CP violation in the lepton sector (for a neutrino physics' overview, see the recent comprehensive book \cite{bib3}). 

Oscillations in vacuum is an interference phenomenon among the matter eigenstates, which is sensitive to their mass-squared differences and to the mixing angles. In the two-flavour framework, the oscillation appearance probability for relativistic neutrinos is given by 
\begin{equation}\label{e:proba}
P(\nu_{\alpha} \rightarrow \nu_{\beta}) = \sin^2(2\theta)\sin^2(\Delta m_{12}^2L/4E), 
\end{equation}
$E$ being the neutrino energy, $L$ the source-detector distance and $\Delta m_{21}^2 = m_{2}^2 - m_{1}^2$. While the oscillation amplitude depends on the mixing angle, the squared-mass differences determine the oscillation frequency\footnote{For three flavours only two independent $\Delta m^2$ can be built. Any extra $\Delta m^2$ requires the addition of more mass eigenstates.}. 
In the last decade, reactor, accelerator and solar experiments have precisely measured the two mass-squared differences to be $\Delta m_{21}^2 = (7.50 \pm 0.20)\, 10^{-5} $eV$^2$,   $|\Delta m_{32}^2| = (2.32 + 0.12) \, 10^{-3}$ eV$^2$, as well as the  mixing angles  $\sin^2(2\theta_{12}) = 0.857 \pm 0.024$,  $sin^2(2\theta_{23}) > 0.95$  \cite{Beringer:1900zz}. An indication for a nonzero third neutrino mixing angle has been found by the T2K \cite{Abe:2011sj} and the Double-Chooz collaborations \cite{Abe:2011fz}; while RENO \cite{Ahn:2012nd} and Daya-Bay have measured $\theta_{13}$ to be $\sin^2(2\theta_{13})=0.092 \pm 0.016(stat) \pm 0.005 (syst) $ \cite{An:2012eh}. 

While the experimental progress is impressive, numerous features remain unrevealed. First, the mechanism that gives a mass to the neutrino is unknown. 
Depending on the neutrino nature, introducing a neutrino mass 
might require a right-handed neutrino singlet (that does not couple to the gauge bosons) or, for example, more complex mechanisms of the see-saw type(s) which need significant  extensions of the Standard Model. There is the mass hierarchy problem. In fact, there are two ways to order the mass eigenstates
since one of the $\Delta m^2$ signs  is unknown. The case $\Delta m^2_{13} > 0$  corresponds to the normal hierarchy, while  $\Delta m^2_{13} < 0$ corresponds to the inverted.   Tritium beta-decay experiments give information on the absolute mass scale. The current upper limit  is of about 2 eV for the electron  neutrino effective mass \cite{Beringer:1900zz}. This will be significantly improved by  the KATRIN experiment that has a discovery potential for a neutrino mass of  0.35 eV at 5 $\sigma$ \cite{Osipowicz:2001sq}. Moreover, key open issues are the existence of leptonic CP violation, of  sterile neutrinos and the identification of the $\nu$ Dirac versus Majorana nature. 

Interestingly,  the ensemble of experimental data present a few anomalies that cannot be cast in the three-active neutrino framework. The neutrino flux measurement from intense	static $^{37}$Ar and $^{51}$Cr sources in the GALLEX and SAGE experiments  present an anomalous deficit of electron neutrinos (the "Gallium anomaly"). A recent analysis finds a  statistical significance at 3 $\sigma$ \cite{Giunti:2010zu}. The LSND collaboration has found evidence for oscillations at $\delta m^2 = 1$ $ eV^2$ (for small mixing angles) using decay-at-rest muons \cite{Athanassopoulos:1996jb} and decay-in-flight pions \cite{Athanassopoulos:1997pv}. Most of the parameter regions identified by LSND have been excluded by the KARMEN experiment based on the same method \cite{Zeitnitz:1998qg}. The MiniBooNE experiment has been built to clarify the controversial LNSD results. However, at present, MiniBooNE anti-neutrino and neutrino oscillation data combined show an excess of events at 3.8 $\sigma$ \cite{AguilarArevalo:2008rc,AguilarArevalo:2012va}. Besides a recent re-evaluation of the electron anti-neutrino flux from reactors has shown a shift in the flux renormalization by 3 $\%$  compared to previous predictions. The re-analysis of all reactor experiments using this new flux has shown a 3 $\sigma$ inconsistency with the standard oscillation framework (the "reactor anomaly") \cite{Mention:2011rk}. These unexplained features might point to new physics and require one or more sterile neutrinos, non-standard interactions or CPT violation. However, at present, none of the proposed explanations is capable of providing a comprehensive understanding. 
\begin{figure}
  \includegraphics[scale=1.]{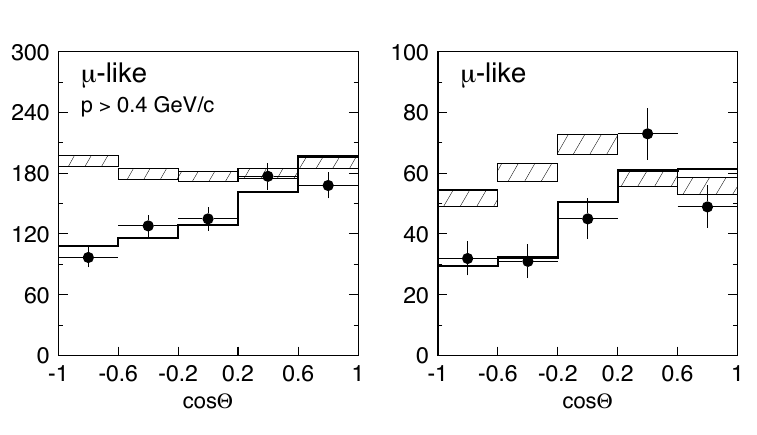}%
  \caption{Zenith angle distribution of (a subset of) the sub-GeV and multi-GeV $\mu$-like events observed by Super-Kamiokande in 1998. Upward-going particles have $\cos \Theta < 0 $ and downward-going particles have $\cos \Theta > 0 $.  The hatched region shows the Monte Carlo expectation for no oscillations normalized to the data live-time with statistical errors. The bold line is the best-fit expectation for  $\nu_{\mu} \rightarrow \nu_{\tau}$ oscillations \cite{Fukuda:1998mi}.}
\end{figure}

Future experiments using man-controlled sources will address unknown neutrino properties and try to identify the possible explanations of the anomalies. Exploring the Dirac CP violating phase requires long-term accelerator measurements employing either established or novel techniques, such as super-beams, beta-beams \cite{Zucchelli:2002sa}, 
neutrino factories  (see e.g. \cite{Volpe:2006in,Bandyopadhyay:2007kx}), or projects like the decay-at-rest based DAEDALUS \cite{Alonso:2010fs}.  
The upgrades of T2K and NO$\nu$A experiments can also investigate a small fraction of the $\delta$ values  at 3 $\sigma$ level \cite{Huber:2009cw}.
Majorana-type phases and the neutrino nature can be determined by searches for the lepton-number violating neutrino-less double-beta decay. Earth matter effects in long-baseline experiments can be exploited to identify the hierarchy, as in e.g. \cite{Bandyopadhyay:2007kx}. Numerous projects are being conceived to test the hypothesis of the existence of sterile neutrinos based e.g. on the use of intense  radioactive sources inside spherical detectors such as $^{144}$Ce in Ref.\cite{Cribier:2011fv} or $^{8}$Li in Ref.\cite{Espinoza:2012jr} (see \cite{Abazajian:2012ys} for a review). 

Astrophysical and cosmological observations offer complementary strategies in these fundamental searches. Cosmological data will reach an unprecedented sensitivity on the sum of the neutrino masses \cite{Lesgourgues:2012uu,Abazajian:2011dt}  and maybe be sensitive to the hierarchy, although this information is indirect. 
Experiments measuring atmospheric neutrinos, like PINGU \cite{Koskinen:2011zz} in IceCube or ORCA \cite{orca} in ANTARES, might reach the required sensitivity 
to determine the mass hierarchy. If an (extra)galactic supernova explosion occurs,  the time and energy signals of supernova neutrinos will have characteristic imprints from
the mass ordering. This would be seen by a network of detectors, such as Super-Kamiokande, KamLAND \cite{Eguchi:2002dm}, Borexino \cite{Arpesella:2008mt},  or by dedicated supernova observatories like HALO \cite{halo,Fuller:1998kb,Engel:2002hg,Vaananen:2011bf} and LVD \cite{Agafonova:2007hn}; while some constitute the "SNEWS: SuperNova Early Warning System" aiming at alerting observers if a galactic supernova occurs \cite{snews}. About a thousand events can be collected  if an explosion happens in our galaxy. Such phenomena are rare (1-3 expected events/century). On the other hand there is the yet unobserved diffuse supernova neutrino background due to supernova explosions
at different cosmological redshifts.  The sensitivity for its discovery could be reached with improved technologies \cite{Beacom:2003nk} or with large-size detectors\footnote{Such detectors are about 20 times larger than Super-Kamiokande. Their goal is to cover an interdisciplinary program including the detection of supernova neutrinos, leptonic CP violation and proton decay searches.} like MEMPHYS  or Hyper-K (440 or 770 kton water Cherenkov) and GLACIER (100 kton liquid Argon) 
\cite{Autiero:2007zj}  that would collect  about 350 and 60 events over 10 years, respectively \cite{Galais:2009wi}. 
From the detection of the diffuse supernova background, key information could be extracted on the supernova dynamics, on the star formation rate and on unknown neutrino properties (for a review see \cite{Beacom:2010kk,Lunardini:2010ab}). 

\section{Neutrino flavour conversion in media : status and open questions}
It is now established that the deficit of high energy solar neutrinos is due to the Mikheev-Smirnov-Wolfenstein (MSW) effect \cite{Wolfenstein:1977ue,Mikheev:1986gs}, a resonant conversion phenomenon occurring when neutrinos interact with the matter composing a medium. 
In two-flavours, the MSW resonance location is identified by the relation\footnote{For antineutrinos the {\it r.h.s.} of the relation has a minus sign.} 
\begin{equation}\label{e:msw}
\sqrt2 G_F n_e = (\Delta m^2 \cos2 \theta)/ 2 E,
\end{equation}
 with $G_F$ the Fermi coupling constant, $n_e$ the electron number density  (see also the early works \cite{Bethe:1986ej, Haxton:1986dm, Bouchez:1986kb}, \cite{bib3} or the recent review \cite{Robertson:2012ib}).
At such a location, a conversion from the electron to the muon (and tau) flavours can take place. Its efficiency (or adiabaticity) depends on the one hand on the mixing angles, on the mass-squared difference values and signs,  and on the matter number density profile on the other (see Figure 3). 
In particular, the evolution is little sensitive to the profile details as far as its smooth enough that the adiabatic condition is met.   
Moreover, depending on the squared-mass difference sign, the MSW effect can occur in the electron neutrino or anti-neutrino channels. 
Since R. Davis' experiment, numerous solar experiments mainly sensitive to the electron flavour have precisely measured the solar neutrino flux. These experiments also had some sensitivity to the other flavours. Using elastic scattering, charged- and neutral-current neutrino interactions on heavy water, 
the SNO experiment has  showed that the measurement of the total $^{8}$B solar neutrino flux is consistent with the predictions of the standard solar model : solar electron neutrinos convert into the other active flavours. In particular, the muon and tau neutrino components of the solar flux has been measured at 5 $\sigma$ \cite{Ahmad:2002jz}. 
Moreover the reactor experiment KamLAND has definitely identified the Large Mixing Angle (LMA) solution, by observing reactor electron anti-neutrino disappearance at an average distance of 200 km. 
The ensemble of these observations shows that averaged vacuum oscillations giving 
\begin{equation}
P (\nu_e \rightarrow \nu_e)  \approx 1 - {1 \over 2} \sin^2 2 \theta_{12} \approx 0.57
\end{equation}
(with $\theta_{12} = 34^{\circ}$) account for the deficit of low energies ($< 2$ MeV) solar neutrinos; 
while the deficit of the high energy portion of the $^{8}$B spectrum is due to the MSW effect. For the latter, the matter-dominated survival probability is 
\begin{equation}
P (\nu_e \rightarrow \nu_e) ^{high~density}  \rightarrow \sin^2\theta_{12} \approx 0.31
\end{equation}
for neutrinos above the critical energy of about 2 MeV Eq.(\ref{e:msw}) (see e.g.\cite{bib3,Robertson:2012ib}). More recently, the Borexino experiment has measured the low energy portion of the solar neutrino spectrum ($pep$ and $^{7}$Be $\nu$)  \cite{Collaboration:2011nga}. 
Figure 4 shows  the results from solar experiments  \cite{Robertson:2012ib}. 
Nowadays the MSW effect constitutes a reference mechanism in the study of neutrino flavour conversion in media. 

\begin{figure}
 \includegraphics[width=8.cm]{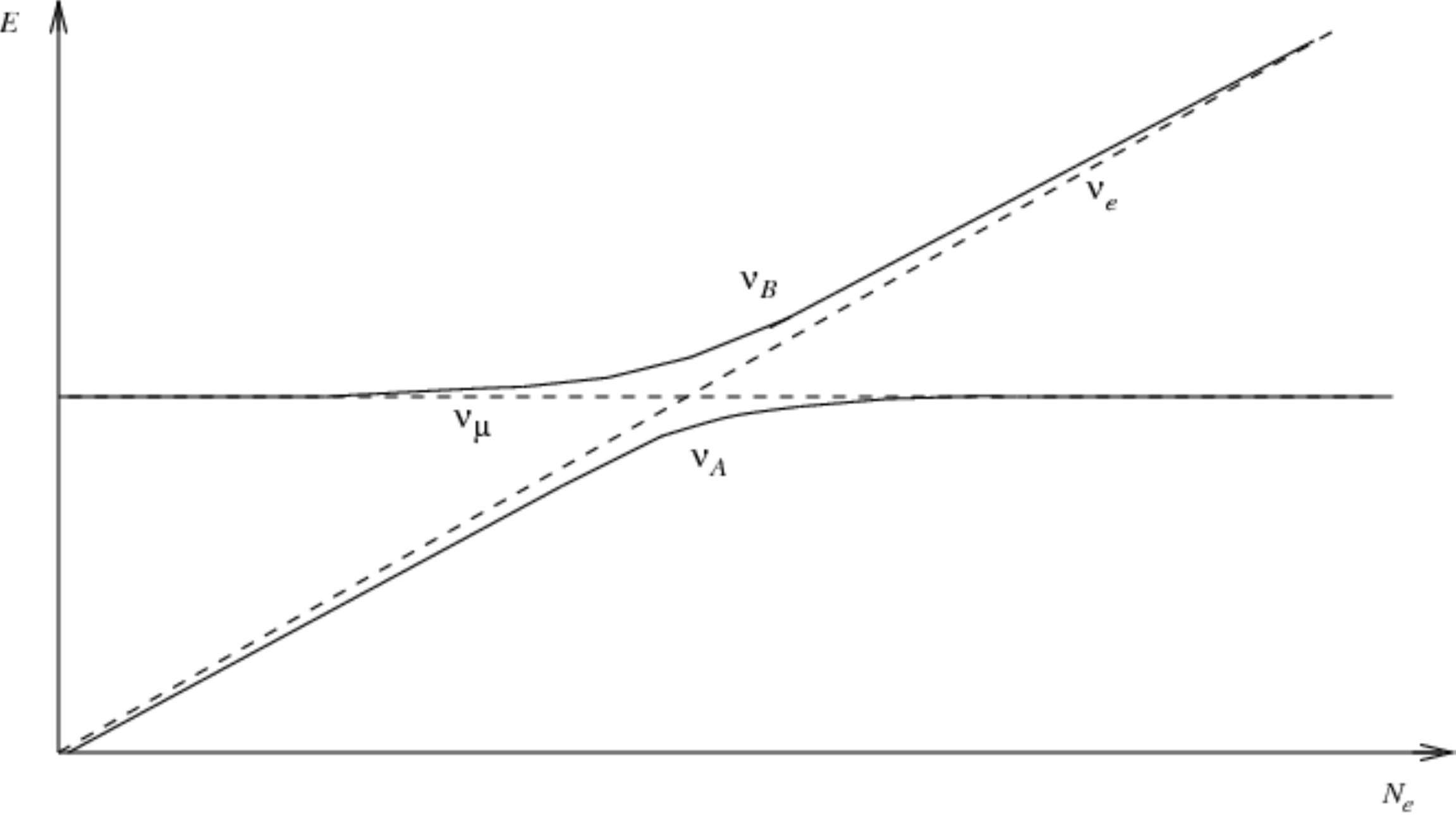}%
 \caption{The MSW effect : Schematic behavior of the neutrino matter eigenvalues (solid lines) as a function of the matter number density of a medium.  The eigenvalues are pushed far apart because of the presence of mixings, at the MSW resonance location. (The dashed lines shows the eigenvalues in absence of mixings.) In the case of an adiabatic evolution the matter eigenstate $\nu_B$ does not mix with the $\nu_A$ at the resonance and stays on the higher branch up to  the surface of the star. In this case an electron neutrino born as $\nu_B$ emerges as  $\nu_B$. This produces an electron neutrino deficit in a $\nu_e$ sensitive detector on Earth. }
\end{figure}
Another interesting case is represented by neutrinos traversing the Earth. Its matter density profile 
undergoes a rapid change from the mantle to the core at higher density \cite{Dziewonski:1981xy}. 
It was first pointed out in \cite{Ermilova:1986ph} and in \cite{akres} that atmospheric neutrinos traversing the Earth would  change in flavour.
Besides the MSW effect, another neutrino flavour conversion mechanism can occur, named "parametric resonance" \cite{Akhmedov:1998ui} or "neutrino oscillation length resonance" \cite{Petcov:1998su}. This effect is an interference effect due to the mantle-core-mantle change in the Earth matter density profile, that can enhance the oscillation probabilities. For example, depending on the trajectory (or azimuthal angle) and energy,
atmospheric neutrinos in the few GeV energy range can have an MSW resonant conversion in the core, or in the mantle and in the core. Neutrinos having core-crossing trajectories can also experience the parametric resonance. The MSW and parametric resonance effects have been investigated e.g. in \cite{Akhmedov:1998xq} in the subdominant atmospheric $\nu_e \rightarrow \nu_{\mu}$ in Super-Kamiokande. An extended literature concerns matter effects in atmospheric neutrinos (see e.g. \cite{Akhmedov:2008qt} and references therein).

\begin{figure}
\begin{center}
 \includegraphics[width=9.cm]{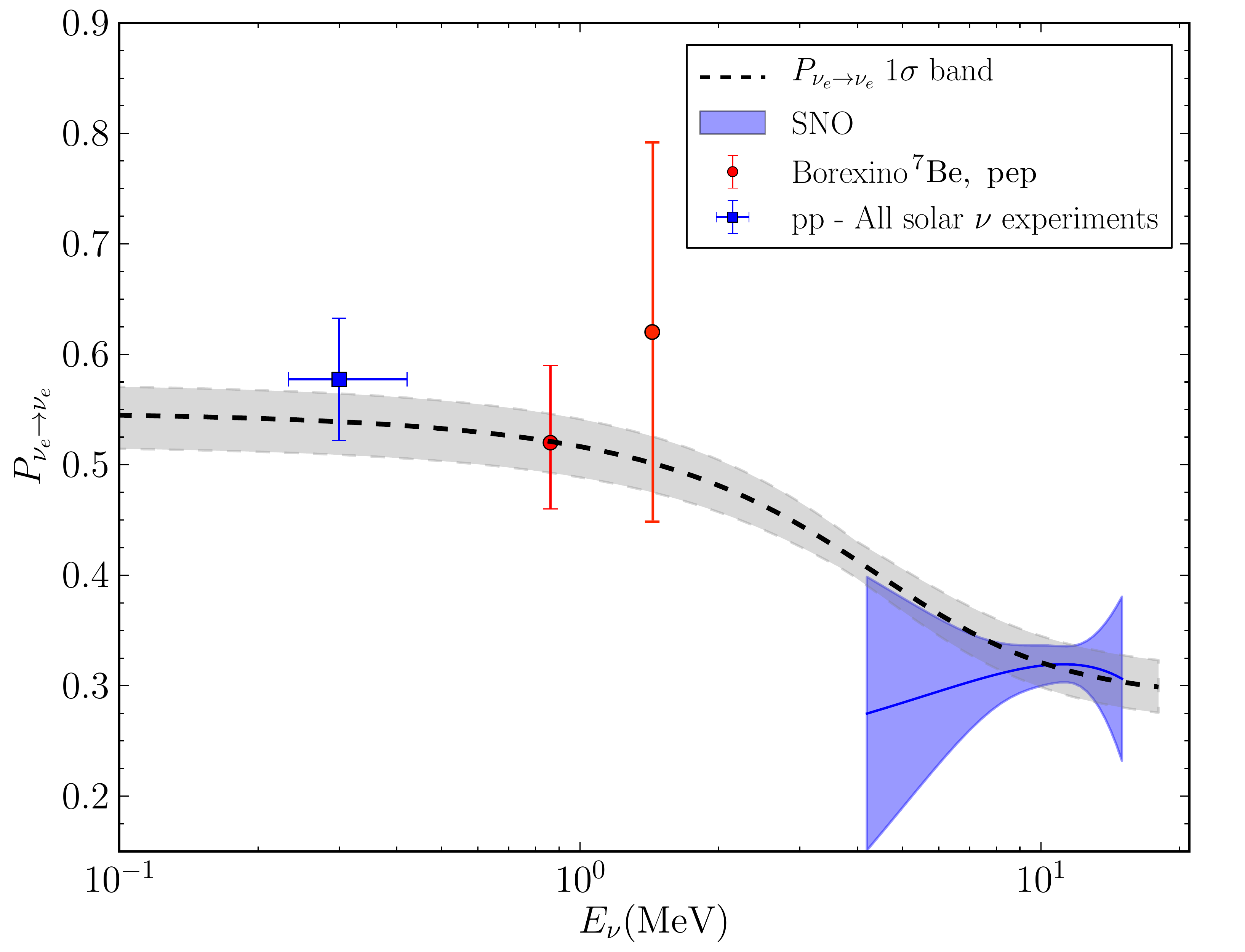}%
\end{center}
  \caption{Solar neutrinos : Electron neutrino survival probability,  as a function of the neutrino energy, for the $pp$, $pep$, $^{7}$Be, $^{8}$B neutrinos from global solar neutrinos analyses, Borexino, and the SNO combined analysis. The results are compared to the MSW predictions, taking into account present uncertainties on mixing angles. The SNO results are from \cite{Aharmim:2011vm} and the ones on $pep$ are from \cite{Collaboration:2011nga}. Figure from \cite{Robertson:2012ib}.}
\end{figure}

A variety of novel flavour conversion mechanisms has been identified in stars with more than 6-8 solar masses - the O-Ne-Mg and iron core-collapse supernovae. Their evolution  ends with an explosion where 99 $\%$ of the gravitational energy is released as neutrinos of all flavours, in the 100 MeV energy range, during a burst lasting about ten seconds. The explosion leaves either a neutron star or a black hole. Since the matter number density of these
stars is very large, the MSW effect occurs at two different locations\footnote{Note that there is also a third resonance named $V_{\mu\tau}$, associated to the $(\theta_{23}, \Delta m^2_{23})$ oscillation parameters. It occurs at higher density than the H-resonance. Since, in general, the $V_{\mu\tau}$ resonance  has a small effect on observations, it will not be discussed further here.}, usually referred to as the High (H) and the Low (L) resonances. The evolution at the H-resonance 
depends on the  $(\theta_{13}, \Delta m^2_{13})$ oscillation parameters, while the flavour evolution at the L-resonance (L) depends on  $(\theta_{12}, \Delta m^2_{12})$ (see Figure 5)  \cite{Dighe:1999bi}. For example, a $40$ MeV neutrino sees the H-resonance at a density of about $10^3$ g/$cm^3$. 
The identification of the solar LMA solution tells us the value and the sign of  $\Delta m^2_{12}$. The three neutrino mixing angles are also now precisely determined. For typical supernova matter density profiles, the neutrino passage in the L-resonance region is adiabatic. On the other hand the neutrino flavour content after the H-resonance encodes interesting information on the mass hierarchy. Therefore matter effects on the supernova neutrino spectra can tell us about the hierarchy from signals of future explosions. However, theoretical investigations of the last ten years have shown that the situation is more complex than what shown in Figure 5 of Ref.\cite{Dighe:1999bi}. This is because one is facing an explosive phenomenon with shock wave(s) and turbulence, with $10^{58}$ MeV released as neutrinos.    
\begin{figure}
 \includegraphics[width=7.cm]{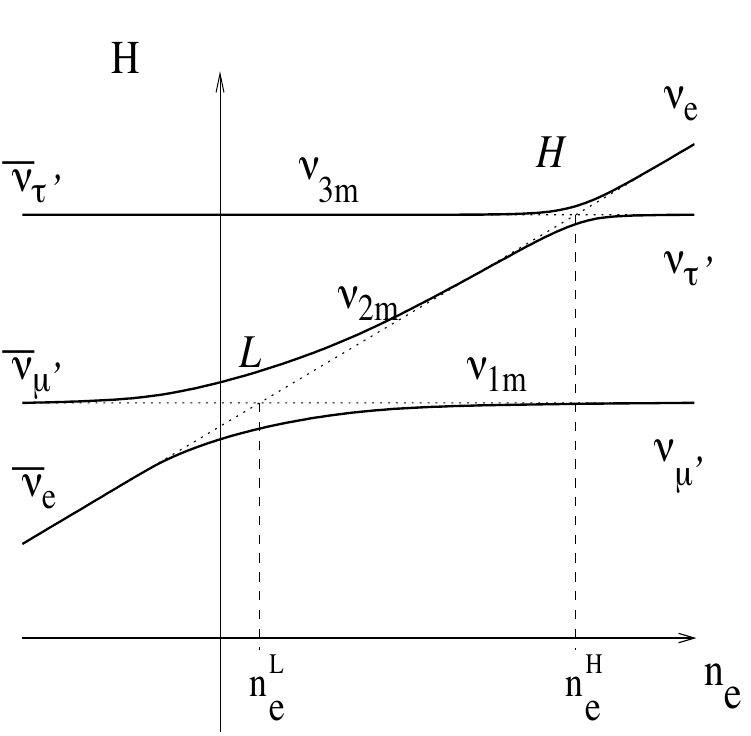}%
 \caption{ MSW effect in supernovae : Level crossing diagram for the neutrino mass eigenstates in a supernova for the case of normal mass hierarchy. The figure shows the  matter eigenvalues evolution associated with the matter eigenstates $\nu_{1m}, \nu_{2m}, \nu_{3m}$, as a function of the electron number density $n_e$. The two crossings correspond to the high-resonance (H) and the low-resonance (L) associated with the mixing parameters $(\theta_{13}, \Delta m^2_{13})$ and ($\theta_{12}, \Delta m^2_{12})$, respectively \cite{Dighe:1999bi}.}
\end{figure}

Ref.\cite{Schirato:2002tg} first pointed out that the shock wave passage in the MSW resonance region would leave an imprint on the time signal of the positrons emitted in inverse beta-decay, i.e. $\bar{\nu}_e + p \rightarrow n + e^+$. This is the main detection channel in Cherenkov and scintillator detectors. 
Ref. \cite{Tomas:2004gr} has emphasized the effects from the presence of not only a front but also a reverse shock, which are apparent in supernova simulations. 
The presence of shock waves engenders two important effects. First it makes the H-resonance temporarily non-adiabatic because of the steepness of the
shock wave fronts (the adiabaticity depends on the derivative of the density profile). Second, multiple H-resonances appear, since
 a neutrino of a given energy can meet the resonance condition several times. For example, in presence of two resonances, the electron neutrino survival probability  reads \cite{Dasgupta:2005wn}
 \begin{equation}
 P_{\nu_e\nu_e} = \cos^2(\chi_1 - \chi_2) - \sin 2\chi_1\sin 2\chi_2 \sin^2(\int_{x2}^{x1} {{\Delta \tilde{m}^2} \over {4E}}) dx 
 \end{equation}
  with $\chi_1$ and  $\chi_2$ the matter angles mixing the two matter eigenstates at the resonance locations $x_1$ and $x_2$, and $\Delta \tilde{m}^2$ the matter squared-mass difference. The interference term oscillates with the neutrino energy and the resonance locations. Since the shock wave is moving, it will change the resonance locations modifying the phase in such a 
term. This produces  rapid oscillations in the survival probability for a neutrino of a given energy. The conditions for such oscillations termed "phase effects"  are that the flavour evolution is semi-adiabatic at the resonances and that the matter eigenstates stay coherent at the resonances  \cite{Dasgupta:2005wn}. 

Turbulence is another characteristic of these explosive environments. Its effects are studied e.g. in \cite{Loreti:1995ae,Fogli:2006xy,Friedland:2006ta}. The presence of matter density fluctuations can introduce numerous locations where the resonance density is met. Their effect on flavour conversion is therefore of the same kind as of shock waves. In fact turbulence produces phase effects, averaging the neutrino oscillation probability between the adiabatic and non-adiabatic solution \cite{Kneller:2010sc} (see also \cite{Lund:2013uta}, and \cite{Duan:2009cd} for a review). Still, extensive calculations  remain to be done, where one implements matter density fluctuations extracted from multi-dimensional supernova simulations (instead of more schematic prescriptions). Clearly, to fully capture several flavour conversion features in media, that are interference effects, one has to evolve the neutrino amplitudes and not the neutrino probabilities. The latter procedure was often followed in the past, e. g. in predictions with the factorisation hypothesis where one factorizes the probabilities for flavour conversion at the H- and L-resonances.  

The implementation of the neutral-current $\nu$-$\nu$ interaction constitutes another important progress in simulations of flavour conversions in supernovae.  Such contributions are small in the sun, whereas the large neutrino number density renders them important  in these environments. Ref. \cite{Pantaleone:1992eq} first pointed out that the neutrino self-interaction would introduce a non-linear refractive index. The first numerical simulations showed the appearance of new phenomena \cite{Samuel:1993uw}. Ref.\cite{Balantekin:2004ug} pointed out a significant impact on the r-process nucleosynthesis. Ref.\cite{Duan:2006an} showed the important effects on the neutrino fluxes with phenomenological implications. This triggered intensive investigations (see \cite{Duan:2010bg} for a review). The inclusion of $\nu$ self-interaction makes
predictions demanding since a system of coupled stiff non-linear equations needs to be solved. 
Three new conversion regimes near the neutrino-sphere are identified.
The first is the "synchronisation"  where neutrino flavour conversion is frozen. The second regime  consists in bipolar oscillations that can
be understood as a "flavour" \cite{Duan:2007mv} or a  "gyroscopic" pendulum \cite{Hannestad:2006nj}. In the last phase, full or no flavour conversion occurs depending on the neutrino energy. The underlying mechanism producing this abrupt change corresponds to an MSW-like behaviour in the co-moving frame \cite{Raffelt:2007cb}, which can be interpreted as a "magnetic resonance" phenomenon  \cite{Galais:2011gh}. This mechanism produces sharp changes of the neutrino fluxes that appear around 200 km from the neutrino-sphere. The electron and muon (or tau) neutrino fluxes swap above a critical energy called the "split" energy. All these effects are also termed "collective" effects because many of their features can be captured by following the neutrino ensemble (instead of a neutrino at a time). These mechanisms explain the novel flavour modifications appearing in simulations with neutrino self-interactions. However the picture becomes more complex depending on the neutrino luminosity ratios at the neutrino-sphere\footnote{The neutrino-sphere is the region in the supernova where neutrinos start free streaming (Figure 6).}, on the neutrino properties and on the detailed implementation of the geometry of the neutrino emission. For example,
large matter densities can decohere collective effects \cite{EstebanPretel:2008ni}. Recent calculations based on realistic matter density profiles from one-dimensional supernova simulations show that indeed such effects might be suppressed during the accretion phase of the supernova explosion \cite{Chakraborty:2011nf}. If so, the situation becomes simpler theoretically because the flavour change is only due to the MSW effect. Nevertheless further modifications can be present due to the shock waves and turbulence, depending on the considered phase of the explosion (neutronization burst, accretion, or cooling phase).  
  
Impressive progress  has been achieved in unravelling mechanisms and conditions for flavour conversion of core-collapse supernova neutrinos. Still, serious work is needed to come to a definite and comprehensive understanding and to establish the impact on observations. 
Current supernova simulations are based on a detailed treatment of the neutrino transport 
in the dense supernova region inside the neutrino-sphere. They typically account for a good angular, or energy neutrino distribution (but not both), and do not include mixings \cite{Janka:2012wk}. Such simulations provide the neutrino energy spectra and fluxes at the neutrino-sphere that are the initial conditions for the flavour evolution studies. These then evolve the $\nu$ up to the star surface in order to predict the associated signals in observatories on Earth (in the case of an explosion). 
The evolution equations used are based on the mean-field approximation. This means that one neutrino (or anti-neutrino) is evolved at a time in a background of matter, of $\nu$ and of $\bar{\nu}$ that acts on the "test" particle through a mean-field. It is still an open question
if and under which conditions the present treatments describe in an appropriate way the transition between the region that is Boltzmann treated to the one that is mean-field described. 
In fact more realistic geometrical descriptions or extended equations implementing many-body corrections might be needed. Works along this line of research are just appearing.
For example,  Ref.\cite{Cherry:2012zw} has pointed out the need for a more realistic geometrical description of the neutrino emission at the neutrino-sphere (Figure 6). 
\begin{figure}
  \includegraphics[width=8.cm]{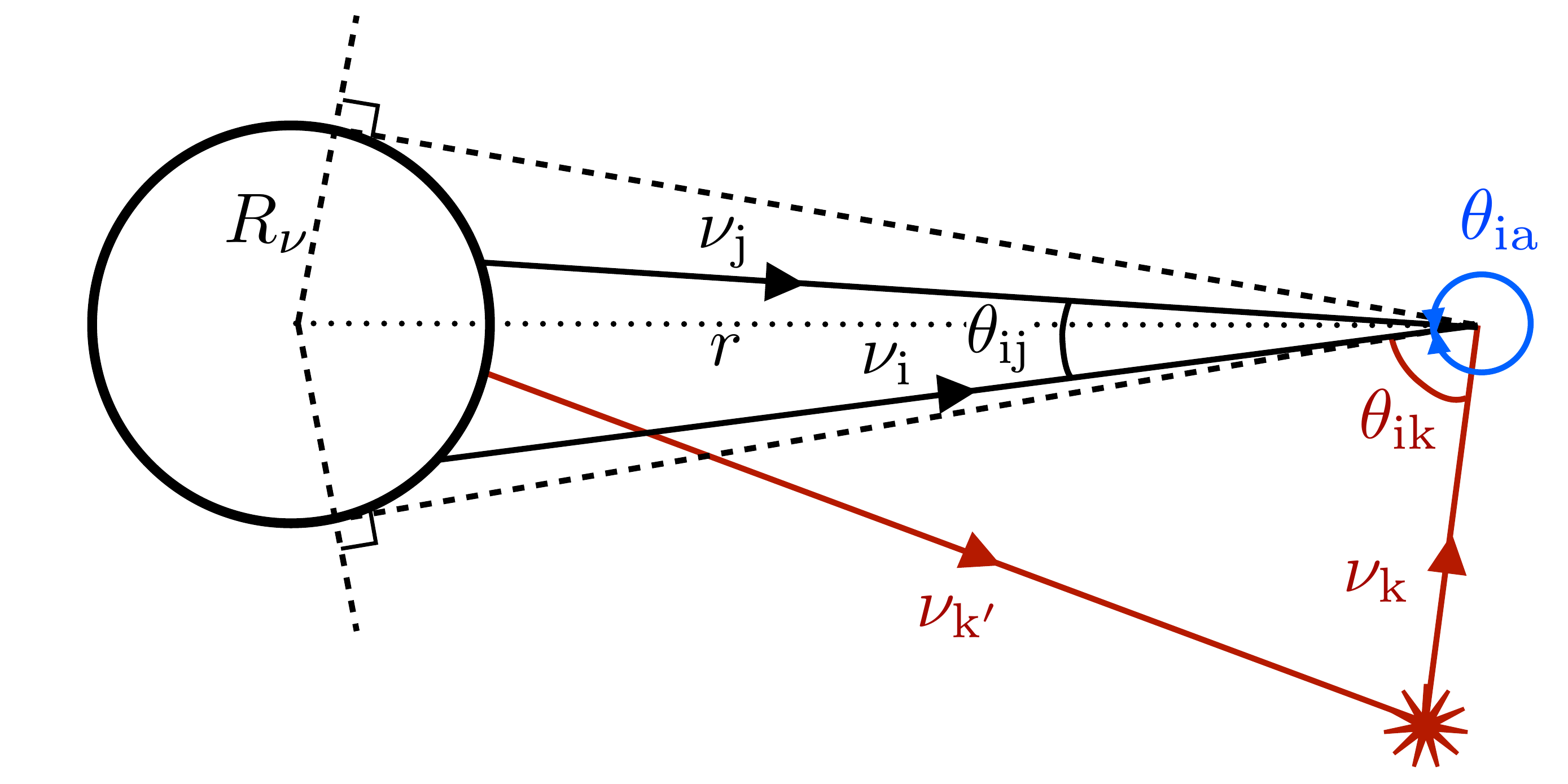}
  \caption{ Cartoon picture of neutrinos emitted at the neutrino-sphere of a supernova and interacting with each other. The neutrino-sphere is taken to be a sharp sphere 
   \cite{Cherry:2012zw}.}
\end{figure} 
Ref.\cite{Balantekin:2006tg} has derived corrections beyond the mean-field approximation using a coherent-state path integral approach. Ref.\cite{Volpe:2013uxl} has used the general many-body framework offered by the  Bogoliubov-Born-Green-Kirkwood-Yvon hierarchy to derive both the mean-field  and extended mean-field equations. The latter  introduce  for the first time the abnormal mean-field corresponding to neutrino-antineutrino pairing correlations (Figure 7). 
So far neglected, these two-body correlations might impact flavour conversion and supernova observations. 

The theoretical developments  described above also apply to the study of the low energy neutrinos from the accretion disks surrounding black holes (AD-BH). In fact the conditions in the disks are very similar to the  core-collapse supernova ones, so that such sites also offer conditions for example for $r$-process and $\nu p$ element nucleosynthesis  (see e.g. \cite{Surman:2005kf,Kizivat:2010ea}). Refs.\cite{Balantekin:2004ug,Malkus:2012ts} have studied e.g. the effect of the neutrino-neutrino interaction in these sites. 

Finally, neutrino flavour conversion phenomena are also important in the Early Universe, in particular at the epoch of Big-bang nucleosynthesis.  Their description is based on the resolution of a Boltzmann equation for particles with mixings since collisions need to be implemented, as well as neutrino mixings, the coupling to the relativistic plasma and neutrino self-interaction. Note that the importance of the neutrino self-interaction was first pointed out in this context.  A key parameter for primordial nucleosynthesis (as for the $r$-process) is the proton-to-neutron ratio.
This is determined by the electron neutrinos and anti-neutrinos interactions with protons and neutrons and by neutron decay.  The build up of the light elements starts at about 1 s after the Big-bang when the plasma temperature is around 1 MeV. The final primordial abundances are very sensitive to neutrino properties (see \cite{Dolgov:2002wy,Iocco:2008va} for a review). Just to give an example, cosmological observations from primordial nucleosynthesis and the matter power spectrum are compatible with about one sterile neutrino (see e.g. \cite{Hamann:2010bk}). The PLANCK experiment has recently measured $N_{eff}= 3.36 \pm 0.66 $ at $95 \% $C.L. (CMB alone) for the effective number of neutrino species \cite{Ade:2013ktc}. The direct observation of the cosmological neutrino background remains a challenge. Its detection demands novel approaches
since such $\nu$s are very cold. 
Attempts are still off from the required sensitivity by several order of magnitudes. The approach which appears as the "closest"  is the detection of cosmological neutrinos
through their capture by radioactive nuclei, a process without threshold. First proposed by Weinberg \cite{Weinberg:1962zza}, this idea has been applied in realistic extensive calculations in \cite{Cocco:2007za} (see also Ref.\cite{Lazauskas:2007da}).
\begin{figure}
  \includegraphics[width=8.cm]{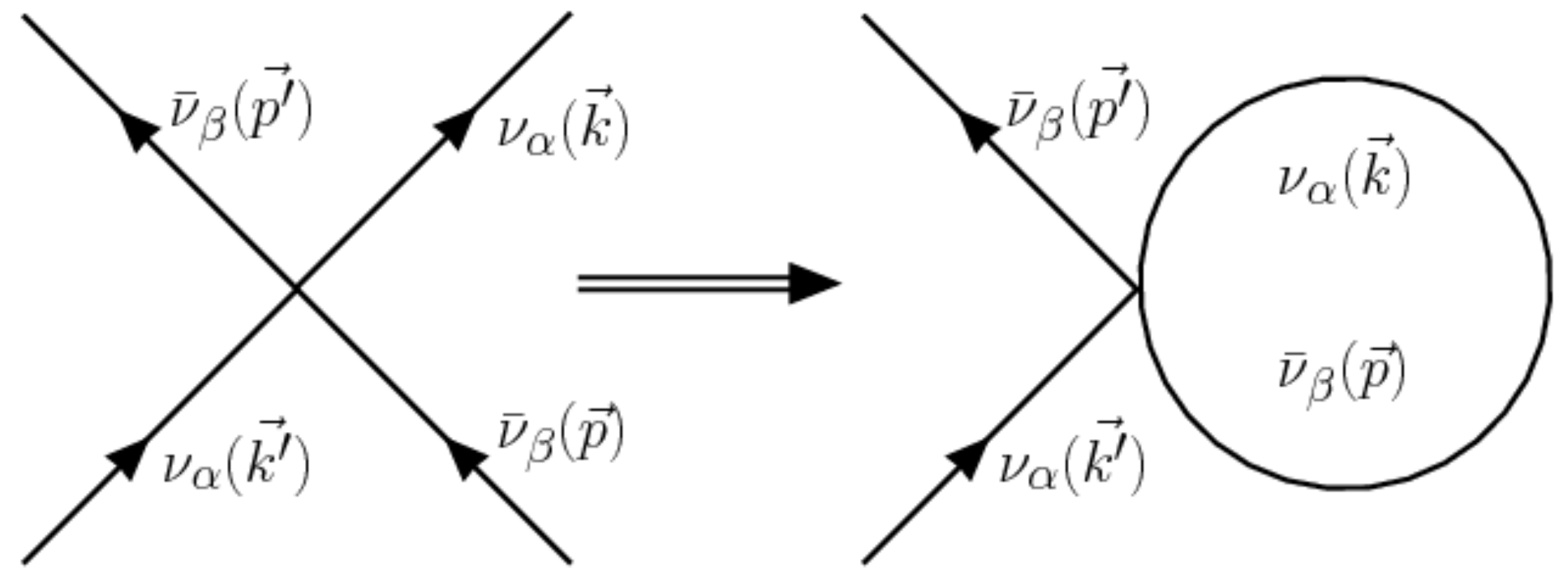}
  \caption{ The (low energy) neutrino-antineutrino interaction (left figure) and the associated pairing mean-field (right figure) obtained by summing over the background particles states. These contributions appear in an extended mean-field description that goes beyond the usual mean-field assumption made in the simulations of flavour conversion in core-collapse supernovae \cite{Volpe:2013uxl}.}
\end{figure}

\section{Open issues and astrophysical $\nu$}
One of the crucial open issues is the possible breaking of CP invariance. The recent precise measurement of the third neutrino mixing angle
by the RENO \cite{Ahn:2012nd} and the Daya-Bay collaborations \cite{An:2012eh} brings a much awaited result for CP searches. In fact CP violation from a non-zero Dirac phase affects oscillations only in the case of three families. The discovery of leptonic CP violation and of neutrino-less double-beta decay would give important clues for the understanding of matter-antimatter asymmetry in the universe (see e.g.\cite{Davidson:2008bu}). An interesting related question to explore is the possibility to have indirect manifestations of a non-zero Dirac-type phase in astrophysical environments.
The authors of Ref.\cite{Minakata:1999ze} have first investigated this question in the solar neutrino context showing that there are no CP violating effects in the $\nu_e$ survival probability in matter to the leading order in electroweak interaction. Effects from next-to-leading order are estimated to be extremely small.
Ref.\cite{Akhmedov:2002zj} has first pointed out possible effects in core-collapse supernovae  coming to the conclusion that there are no CP violation effects in such environments. 
Ref.\cite{Balantekin:2007es} has come to a result at variance with Ref.\cite{Akhmedov:2002zj}  demonstrating that there can be CP violating effects in supernovae.
By using general arguments, the factorisation condition 
\begin{equation}
H( \delta ) = S^{\dagger} H ( \delta = 0 ) S
\end{equation}
 is obtained\footnote{$H$ being in the so-called $T_{23}$ basis}  with $S = diag(1,1,e^{i\delta})$.  This relation gives a procedure to identify under which conditions such effects can arise : contributions to the neutrino Hamiltonian  that break the factorization condition engender CP violating effects. For example, these arise because of radiative corrections in the Standard Model, or of non-standard interactions like flavour-changing interactions that differentiate muon from tau neutrino interactions with matter. This finding has been subsequently independently confirmed by Ref.\cite{Kneller:2009vd}.
A quantitative evaluation of the modifications introduced by a non-zero Dirac phase shows variations at the level of a few percent both of the supernova signal in an observatory on Earth or of the neutron-to-proton ratio relevant for $r$-process nucleosynthesis \cite{Balantekin:2007es}. 
An improved numerical simulation with the neutrino self-interaction has shown that the non-linearity of the equations amplifies the CP violating effects from radiative corrections by several order of magnitudes \cite{Gava:2008rp}. While flux modifications are still at the level of about 5-10 $\%$, future improvements of the simulations might introduce further amplification of such effects. Ref.\cite{deGouvea:2013zp} has studied the size of CP violating effects in presence of non-zero neutrino magnetic moments, reaching the same conclusions as previous works  \cite{Balantekin:2007es,Gava:2008rp} concerning the size of CP effects.
In the Early universe context Ref.\cite{Gava:2010kz} has performed the first investigation of the effects of a non-zero Dirac-type phase on the primordial light elements abundances. Within three-active neutrinos it is shown that the phase modifies the primordial Helium-4 abundance at most by $1 \%$, which is within current systematic uncertainties.

\begin{figure}
   \includegraphics[width=7.cm,angle=-90]{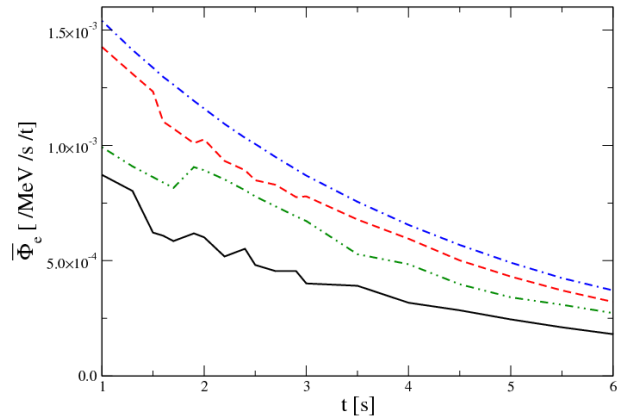}
  \caption{ Positron time signal associated with inverse beta-decay 
for a galactic explosion at 10 kpc per unit tonne  in a detector. The different curves are for 
10 (solid), 15 (dashed), 19 (dash-dotted) and 29 
(dot-dot-dashed line) MeV positron energies. The dips or bumps correspond to the passage of the
shock wave in the MSW resonance region. They are present in the case of inverted mass hierarchy \cite{Gava:2009pj}.}
\end{figure}
The detection of neutrinos from a future (extra)galactic explosion constitutes one of the strategies to identify the mass hierarchy. Available studies are roughly of three types. The first kind exploits Earth matter effects of supernova neutrinos that traverse the Earth before being detected  \cite{Dighe:1999bi}. This produces a modulation of the neutrino events distribution that gives peaks in the distribution Fourier transform \cite{Dighe:2003be}. However such effects are very small making this option hardly feasible \cite{Borriello:2012zc}.
A second option consists in exploiting the early time signal of the supernova explosion either from the neutronization burst \cite{Kachelriess:2004ds} or from the accretion phase.
 In Ref.\cite{Serpico:2011ir} the signal of the first 200 ms (accretion phase) of the explosion is investigated. This time window has the advantage of being simpler than the late explosion stages since the signal is not affected by the neutrino-self-interaction, the shock-waves (and turbulence). 
The authors show that in IceCube one could distinguish the normal from the inverted hierarchy. 
The third type of studies look for effects in the time signal due to the shock waves since their presence depends on the hierarchy \cite{Tomas:2004gr, Barger:2005it}. 
Early works show features without the $\nu-\nu$ interaction. This improvement is performed in Ref.\cite{Gava:2009pj}. The calculation treats neutrino self-interactions and shock-wave effects in a consistent way, using realistic supernova density profiles and propagating neutrino amplitudes (phase effects are properly taken into account). 
Figure 8 shows the results for the positron time signal associated with inverse beta-decay in water Cherenkov and scintillator detectors. At early times, when the shock wave has not yet reached the MSW resonance region, the neutrino conversion is expected to be adiabatic. As the shock wave reaches this region, the flux becomes the non-adiabatic one, producing either a dip or a bump, depending on the neutrino energy. For electron anti-neutrinos, the resonance condition is met for inverted hierarchy. If the hierarchy is normal, no resonant conversion occurs and the positron time signal presents an exponential behaviour. Such a signature could already be seen in Super-Kamiokande if a galactic supernova happens tomorrow, and be distinguished by the exponential at 3.5 $\sigma$  (1 $\sigma$) for the bump (dip). 
Note that this signature holds in the electron neutrino channel as well, if the mass hierarchy is normal. This could be seen in a liquid argon detector mainly sensitive to $\nu_e$ or  in water Cherenkov and in scintillator detectors through  scattering on oxygen and carbon respectively. 
 
Complementary information would be obtained by the observation of (extra)galactic supernova explosion(s) and the detection of the diffuse supernova neutrino background. 
While some of the uncertainties are still large, one can pin down interesting information by combining 
 data from different detectors.  Having different technologies sensitive to $\nu$ and $\bar{\nu}$ with various energy thresholds is a key aspect to be sensitive to different parts of the neutrino spectra. Figure 9 shows the example 
 of the interest of having one- and two-neutron detection channels in a lead-based detector like HALO \cite{Vaananen:2011bf}. Note that a software - SNOwGLoBES - is now available  to compute interaction rates for supernova neutrinos in common detector materials \cite{snowglobes,Scholberg:2012id}. 

Observations using the $\nu_e$ detection channel would benefit from precise measurements of neutrino-nucleus scattering. Only for deuteron the cross section is known at the level of few percent, while for heavier nuclei  uncertainties  are typically at the level of few tens of percent. In the future, facilities producing neutrinos in the 100 MeV energy range such as low energy beta-beams \cite{Volpe:2003fi} or spallation sources \cite{Avignone:2003ep}, could offer a unique opportunity to precisely measure such cross sections and also realise fundamental interaction studies, like measurements of the Weinberg angle at low momentum transfer, a test of the Conserved-Vector-Current hypothesis  (see \cite{Volpe:2006in} for a review), or of the hypothesis of the existence of sterile neutrinos \cite{Espinoza:2012jr}. Note that improving the knowledge of the isospin and spin-isospin nuclear response involved in neutrino-nucleus interactions is of importance  also for searches on the neutrino nature, since this constitutes a key ingredient of neutrino-less double-beta decay predictions 
\cite{Volpe:2005iy}.  
\begin{figure}
  \includegraphics[width=9.cm]{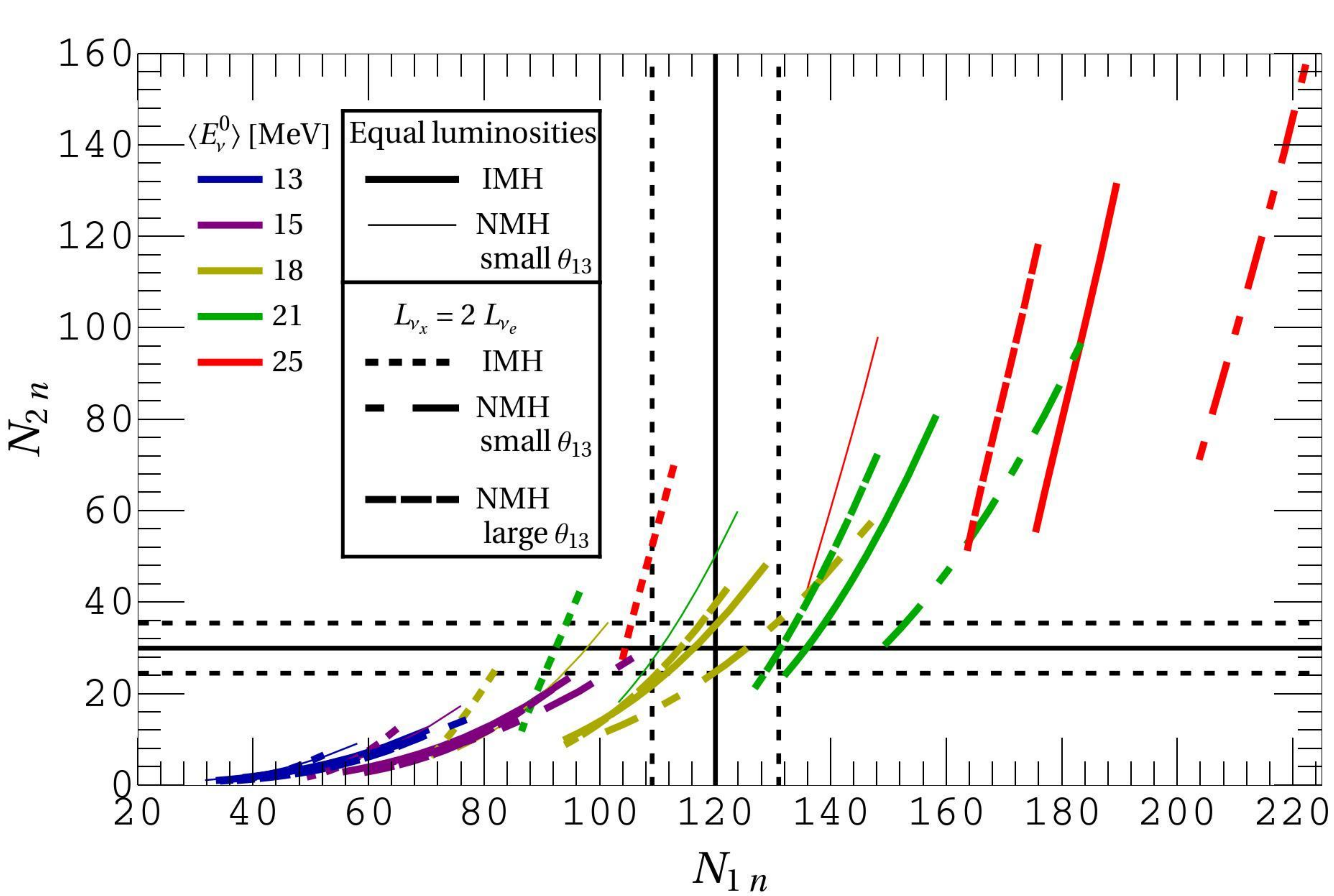}%
  \caption{  One- and two-neutron emission event rates for a supernova at 10 kpc in the HALO phase-II detector (1 kton of $^{208}$Pb). The different curves take into account the uncertainties in the supernova muon (tau) neutrino fluxes at the neutrino-sphere as well as the possible hierarchy ($\theta_{13}$ is now measured)
  \cite{Vaananen:2011bf}.}
\end{figure}

 Observing a supernova luminosity curve is of great astrophysical value besides being of interest for the fundamental particle or interaction properties it encodes. In fact, 
the $\nu$ time signal closely follows  the different phases of the explosion, from the collapse to the accretion phase and to the cooling of the remaining proto-neutron star. This measurement would provide key information for the longstanding open problem of the explosion mechanism of iron-core supernovae.
 Currently, 2- and 3-dimensional simulations are being developed that realistically include the neutrino transport, convection and hydrodynamical instabilities (the Standing-Accretion-Shock o SASI mode). Several groups are obtaining successful explosions for different progenitor masses, while reaching a consensus on the mechanism 
requires further studies (see e.g. \cite{Janka:2012wk,Mezzacappa:2005ju,Kotake:2005zn}). Using 3D models based on
a simplified neutrino transport scheme, Ref.\cite{Lund:2012vm} has showed for example that the SASI mode could be tracked in IceCube for a supernova within 2 kpc. These massive stars are also candidate sites for the $r$-process, whose identification is still an open issue; while several sites might contribute, such as the accretion disks around black holes   
  \cite{Abe:2011fz}. The impact of neutrino properties on the $r$-process is the object of numerous studies. For example Ref.\cite{McLaughlin:1999pd} points out the possible role of sterile neutrinos in getting a successful nucleosynthesis in supernovae. An intriguing question is also the possible impact of flavour conversion on the explosion itself. MSW effects occur in the outer star layers and have no impact on the shock. The collective effects from the neutrino-neutrino interaction take place deep in the star and might have an impact. The first investigations seem to indicate that they are still far out to affect the shock waves \cite{Dasgupta:2011jf}, although it might still be too early to draw conclusions.

A new window on the universe is opened by high-energy neutrino telescopes like ANTARES, the first undersea in the Mediterranean   \cite{Collaboration:2011nsa}, and IceCube, buried in deep ice at the South Pole \cite{Kappes:2012hk}. The main mission is to search for galactic and extra-galactic sources of high-energy neutrinos to elucidate the source of cosmic rays and the astrophysical mechanisms that produce them. These telescopes also investigate neutrino oscillations, dark matter and supernova neutrinos (for IceCube). At present, the IceCube collected data on ultra-high-energy neutrinos (above $10^{19}$ eV) from active galactic nuclei (AGN) and gamma-ray-bursts (GRB) show no significant deviation from background in the number of observed events, although uncertainties in the models are still quite large.
Interestingly two 1 PeV events at threshold of GZK searches have been recently identified  with a significance of  2.8 $\sigma$ of not being a background  \cite{Kappes:2012hk}. 
A follow-up analysis of contained vertex event search has just found 26 more events at lower energy (3.6 $\sigma$). There is now evidence for a high-energy neutrino component inconsistent at 4.3 $\sigma$ with standard atmospheric backgrounds \cite{HE-icecube}. A future larger data set with increased statistical significance might confirm the discovery of high energy astrophysical neutrinos. 

High-energy neutrino telescopes are currently also providing data on neutrino oscillations measuring atmospheric neutrinos, commonly a background for astrophysical neutrino searches.   
Using low energy samples, both ANTARES \cite{AdrianMartinez:2012ph} and IceCube/DeepCore \cite{Gross:2013iq} have measured the parameters $\theta_{23}$ and $ \Delta m^2_{23}$ in good agreement with existing data. 
Crucial information can be extracted from such telescopes in the future from low energy (below 40 GeV) atmospheric neutrinos. Predictions show that from the measurement of flavour conversion effects in the Earth in  IceCube/DeepCore can provide a precise measurement of oscillation parameters (see e.g. \cite{FernandezMartinez:2010am,Mena:2008rh}). PINGU, IceCube extension in the 10 GeV energy range, could measure the mass hierarchy and be sensitive to the Dirac phase \cite{Akhmedov:2012ah}. Feasibility studies are currently ongoing both for PINGU and for ORCA to establish if the energy and angular resolution required for the mass hierarchy search can be achieved.
Information on the phase could also be extracted by a precise measurement of high-energy neutrino flux ratios \cite{Serpico:2005sz,Winter:2006ce,Meloni:2012nk}.
Ref.\cite{Meloni:2012nk} gets a 2 $\sigma$ coverage of about 10 $\%$ of the CP values, if uncertainties on the flux ratios are kept below 7 $\%$. As for a fourth sterile neutrino, its existence might lead to a distortion of the zenith angle
distribution of high-energy atmospheric neutrinos through the MSW active-sterile resonance inside the Earth. 
An analysis of data from Amanda and (incomplete) IceCube  as well as the prospects for the completed IceCube give limits that for some of the sterile oscillation parameters are stronger than all combined experiments \cite{Esmaili:2012nz}. Ref. \cite{Razzaque:2012tp} shows that low energy events in DeepCore can substantially constrain the mixing of sterile neutrinos in the eV mass scale. The effect found is different for normal or inverted hierarchy of active neutrinos, so that if such $\nu_s$s exist the hierarchy can also be identified.  
Finally both terrestrial experiments and neutrino telescopes  \cite{Abbasi:2010kx} set tight limits on Lorentz and CPT violation. A constraint also comes from the SN1987A from the nearly simultaneous arrival of the photons and the neutrinos \cite{Longo:1987ub}.
Interestingly,  models based on Lorentz and CPT violation are developed to interpret oscillation data from all experiments  (see \cite{Katori:2012hc} and the nice summary in Ref.\cite{Diaz:2011tx}). 

In conclusion, neutrinos are messengers having a wide energy range, capable of covering cosmological distances and telling us about quiet and violent phenomena.
Neutrino physics is a domain rich of interdisciplinary aspects and approaches. The measurement of these elusive particles
requires inventive detectors of huge sizes. Sometimes slow on a man-scale, the progress continues steadily, while the last decade has been rich of exciting observations. Neutrinos have brought milestones in our understanding of fundamental issues in high-energy physics and astrophysics, and will likely bring more in the future.


\begin{thebibliography}{99}

  \bibitem{Davis:1964hf}%
  \textsc{R.~Davis}, 
  Phys.\ Rev.\ Lett.   \textbf{12}, 303 (1964).

\bibitem{Hirata:1989zj} 
   \textsc{K.~S.~Hirata et al.},
  J Phys.\ Rev.\ Lett. \textbf{63}, 16 (1989).


\bibitem{Abdurashitov:1999zd} 
 \textsc{  J.~N.~Abdurashitov et al.},
J  Phys.\ Rev.\ C \textbf{60}, 055801 (1999).


\bibitem{Anselmann:1994cf} 
  \textsc{P.~Anselmannet al.}  
  Phys.\ Lett.\ BÊ \textbf{327}, 377 (1994).

\bibitem{Arpesella:2008mt} 
  \textsc{  C.~Arpesella et al.},
    Phys.\ Rev.\ Lett. \textbf{101}, 091302 (2008).

 \bibitem{Ahmad:2002jz}%
 \textsc{Q.~R.~Ahmad et al.},
  Phys.\ Rev.\ Lett. \textbf{89}, 011301 (2002).   
  






\bibitem{Bahcall:1987jc} 
  \textsc{   J.~N.~Bahcall},  and  
   \textsc{ R.~K.~Ulrich},
  Rev.\ Mod.\ Phys. \textbf{60}, 297 (1988).

\bibitem{TurckChieze:1993dw} 
 \textsc{   S.~Turck-Chieze }, and 
 \textsc{  I.~Lopes},
 Astrophys.\ J. \textbf{408}, 347 (1993).


\bibitem{Adelberger:1998qm} 
  \textsc{  E.~G.~Adelberger et al.},
    Rev.\ Mod.\ Phys. \textbf{70}, 1265 (1998).


\bibitem{Robertson:2012ib} 
   \textsc{ W.~C.~Haxton}, 
    \textsc{ R.~G.~Hamish Robertson}, and 
     \textsc{ A.~M.~Serenelli},
  arXiv:1208.5723.

   \bibitem{Suzuki:2008zzf}%
  \textsc{A.~Suzuki}, 
 J.\ Phys.\ Conf.\ Ser. \textbf{120}, 072001 (2008).
  
    \bibitem{Hirata:1987hu}%
  \textsc{K.~Hirata et al.}, 
  Phys.\ Rev.\ Lett. \textbf{58}, 1490 (1987).
  
      \bibitem{Bionta:1987qt}%
  \textsc{R.~M.~ Bionta et al.}, 
   Phys.\ Rev.\ Lett. \textbf{58}, 1494 (1987).
 
 
 \bibitem{Alexeyev:1987}%
  \textsc{Alexeyev, E.N. et al. }, 
Proc. of the Leptonic Session of the 22nd Rencontre de Moriond (1987) 739.
  
  
 \bibitem{Aglietta:1987it}%
  \textsc{M.~Aglietta et al. }, 
  Europhys.\ Lett. \textbf{3}, 1315 (1987).
  
  
  \bibitem{Pagliaroli:2008ur}%
  \textsc{G.~Pagliaroli}, 
 \textsc{F.~Vissani}, 
 \textsc{M.~L.~Costantini},  and
 \textsc{A.~Ianni}, 
Astropart.\ Phys. \textbf{31},  163 (2009).

\bibitem{Hirata:1988uy} 
   \textsc{K.~S.~Hirata et al.},
  Phys.\ Lett.\ B \textbf{ 205}, 416 (1988).

  \bibitem{Fukuda:1994mc} 
   \textsc{  Y.~Fukuda  et al.},
   Phys.\ Lett.\ B  \textbf{  335}, 237 (1994).
  
  
\bibitem{Casper:1990ac} 
  \textsc{  D.~Casper et al.},
  Phys.\ Rev.\ Lett. \textbf{  66}, 2561 (1991).
  
 \bibitem{Berger:1990rd} 
   \textsc{  C.~Berger   et al.}, 
Phys.\ Lett.\ B \textbf{ 245}, 305 (1990).
  
  \bibitem{Aglietta:1988be} 
   \textsc{   M.~Aglietta et al.},
    Europhys.\ Lett. \textbf{8}, 611 (1989).
  
\bibitem{Fogli:1996nn} 
\textsc{ G.~L.~Fogli},
\textsc{ E.~Lisi}, 
\textsc{D.~Montanino}, and   \textsc{G.~Scioscia},
Phys.\ Rev.\ D  \textbf{55}, 4385 (1997).
  
 \bibitem{Fukuda:1998mi}%
  \textsc{Y.~Fukuda et al.}, 
   Phys.\ Rev.\ Lett. \textbf{81}, 1562 (1998).

\bibitem{Bethe:1939bt}
  \textsc{H.A. Bethe},
  Phys. Rev.   \textbf{55}, 1562 (1939).
  
\bibitem{gs}
  \textsc{G. Gamow}, and
 \textsc{M. Schoenberg},  
  Phys. Rev.   \textbf{58},  1117 (1940).


 \bibitem{Pontecorvo:1957cp}%
  \textsc{ B.~Pontecorvo,},
   Sov.\ Phys.\ JETP  \textbf{6},  429 (1957)
  [Zh.\ Eksp.\ Teor.\ Fiz.  \textbf{33}, 549 (1957)].

\bibitem{Maki:1962mu}%
\textsc{Z.~Maki},
  \textsc{M.~Nakagawa},  and
   \textsc{S.~Sakata}, 
 Prog.\ Theor.\ Phys. \textbf{28}, 870 (1962).


 
   \bibitem{Beringer:1900zz}%
  \textsc{J.~Beringer et al.}, 
   Phys.\ Rev.\ D \textbf{86}, 010001 (2012).


    \bibitem{bib3}%
  \textsc{C.~Giunti}, 	and
  \textsc{C.~W.~Kim}, 
 ``Fundamentals of Neutrino Physics and Astrophysics,'' (Univ. Pr., Oxford, 2007), p.\,710.

 \bibitem{Abe:2011sj}%
  \textsc{K.~Abe et al.}, 
Phys.\ Rev.\ Lett. \textbf{107}, 041801 (2011).

 \bibitem{Abe:2011fz}%
  \textsc{Y.~Abe et al.}, 
Phys.\ Rev.\ Lett. \textbf{108}, 131801 (2012).


 \bibitem{Ahn:2012nd}%
  \textsc{J.~K.~Ahn et al.}, 
arXiv:1204.0626.


\bibitem{An:2012eh}
  \textsc{F.~P.~An et al.}, 
  Phys.\ Rev.\ Lett. \textbf{108}, 171803 (2012).

\bibitem{Osipowicz:2001sq}%
  \textsc{A.~Osipowicz et al.}, 
hep-ex/0109033.

 \bibitem{Giunti:2010zu}%
  \textsc{C. Giunti} and
  \textsc{M. Laveder},
  Phys. Rev. C \textbf{83}, 065504 (2011).

 \bibitem{Athanassopoulos:1996jb}%
  \textsc{C. Athanassopoulos et al.}, 
  Phys. Rev. Lett. \textbf{77}, 3082 (1996).

\bibitem{Athanassopoulos:1997pv} 
   \textsc{C.~Athanassopoulos et al.},
    Phys.\ Rev.\ Lett. \textbf{ 81}, 1774 (1998).

\bibitem{Zeitnitz:1998qg} 
    \textsc{B.~Zeitnitz et al.},
    Prog.\ Part.\ Nucl.\ Phys. \textbf{ 40}, 169 (1998).

 \bibitem{AguilarArevalo:2008rc}%
  \textsc{A. A. Aguilar-Arevalo et al.}, 
  Phys. Rev. Lett. \textbf{102},  (2009) 101802.

  
 \bibitem{AguilarArevalo:2012va}%
  \textsc{ A.~A.~Aguilar-Arevalo et al.}, 
  arXiv:1207.4809 [hep-ex].
 

 \bibitem{Mention:2011rk}%
  \textsc{G. Mention et al.}, 
  Phys. Rev. D  \textbf{83}, 073006 (2011).


\bibitem{Zucchelli:2002sa} 
 \textsc{ P.~Zucchelli},
   Phys.\ Lett.\ B \textbf{ 532}, 166 (2002).

 \bibitem{Volpe:2006in}%
  \textsc{C.~Volpe}, 
  J.\ Phys.\ G \textbf{34}, R1 (2007).
  
  
  
  \bibitem{Bandyopadhyay:2007kx} 
  \textsc{A.~Bandyopadhyay et al.}, 
    Rept.\ Prog.\ Phys. \textbf{72}, 106201 (2009).


 \bibitem{Alonso:2010fs}%
  \textsc{J.~Alonso et al.}, 
  arXiv:1006.0260.
  
  
  \bibitem{Huber:2009cw} 
  \textsc{P.~Huber et al.}, 
   JHEP  \textbf{ 0911}, 044 (2009).



 \bibitem{Cribier:2011fv}%
  \textsc{M.~Cribier et al.}, 
  Phys. Rev. Lett. \textbf{83},   201801 (2011).

 \bibitem{Espinoza:2012jr}%
  \textsc{C.~Espinoza}, 
  \textsc{R.~Lazauskas}, and 
  \textsc{C.~Volpe},
  Phys.\ Rev.\ D \textbf{1986}, , 113016 (2012).

 \bibitem{Abazajian:2012ys}%
  \textsc{K.~N.~Abazajian et al.}, 
  arXiv:1204.5379.


 \bibitem{Lesgourgues:2012uu}%
  \textsc{J.~Lesgourgues}  and 
  \textsc{S.~Pastor},
  Adv.\ High Energy Phys. \textbf{2012}, 608515 (2012).
 
  \bibitem{Abazajian:2011dt}%
  \textsc{K.~N.~Abazajian et al.}, 
Astropart.\ Phys. \textbf{35}, 177 (2011).
  
  
   \bibitem{Koskinen:2011zz}%
  \textsc{D.~J.~Koskinen}, 
Mod.\ Phys.\ Lett.\ A \textbf{26}, 2899 (2011).
  
  \bibitem{orca}%
 see http://agenda.infn.it/conferenceDisplay.py?\\confId=5510 
  
     \bibitem{Eguchi:2002dm}%
 \textsc{K.~Eguchi  et al.}, 
 Phys.\ Rev.\ Lett. \textbf{90}, 021802 (2003).   

  
    \bibitem{halo}%
see http://www.snolab.ca/halo/  

    
  
\bibitem{Fuller:1998kb} 
  \textsc{  G.~M.~Fuller},   \textsc{W.~C.~Haxton},
   and   \textsc{G.~C.~McLaughlin},
   Phys.\ Rev.\ D \textbf{59}, 085005 (1999).
  
  
  \bibitem{Engel:2002hg} 
    \textsc{J.~Engel},   \textsc{G.~C.~McLaughlin}, and   \textsc{C.~Volpe},
   Phys.\ Rev.\ D \textbf{ 67}, 013005 (2003).

  
   \bibitem{Vaananen:2011bf}%
  \textsc{D.~V\"a\"an\"anen}, and
  \textsc{  C.~Volpe},
JCAP \textbf{1110}, 019 (2011).


  
\bibitem{Agafonova:2007hn} 
  \textsc{N.~Y.~.Agafonova t et al.},
 J Astropart.\ Phys.  {\bf 28}, 516 (2008).
 
  \bibitem{snews}%
see http://snews.bnl.gov/  
  

 \bibitem{Beacom:2003nk}%
  \textsc{J.~F.~Beacom}, and 
  \textsc{M.~R.~Vagins},
  Phys.\ Rev.\ Lett. \textbf{93}, 171101 (2004).


\bibitem{Autiero:2007zj} 
  \textsc{D.~Autiero et al.}, 
JCAP  \textbf{0711}, 011 (2007).


 \bibitem{Galais:2009wi}%
  \textsc{S.~Galais}, 
  \textsc{J.~Kneller}, \textsc{C.~Volpe},
  and 
  \textsc{J.~Gava},
  Phys.\ Rev.\ D \textbf{81}, 053002 (2010).


\bibitem{Beacom:2010kk}%
  \textsc{J.~F.~Beacom}, 
  Ann.\ Rev.\ Nucl.\ Part.\ Sci. \textbf{60}, 439 (2010).


 \bibitem{Lunardini:2010ab}%
  \textsc{C.~Lunardini}, 
  arXiv:1007.3252.




  \bibitem{Wolfenstein:1977ue}%
  \textsc{L.~Wolfenstein}, 
  Phys.\ Rev.\ D \textbf{17}, 2369 (1978).

  \bibitem{Mikheev:1986gs}%
  \textsc{S.~P.~Mikheev}, and 
  \textsc{A.~Y.~Smirnov},
  Sov.\ J.\ Nucl.\ Phys. \textbf{42}, 913 (1985).



\bibitem{Bethe:1986ej} 
 \textsc{   H.~A.~Bethe},
    Phys.\ Rev.\ Lett. \textbf{ 56}, 1305 (1986).

\bibitem{Haxton:1986dm} 
   \textsc{ W.~C.~Haxton},
    Phys.\ Rev.\ Lett. \textbf{57}, 1271 (1986).

\bibitem{Bouchez:1986kb} 
 \textsc{   J.~Bouchez et al.},
    Z.\ Phys.\ C \textbf{32}, 499 (1986).

\bibitem{Aharmim:2011vm} 
  \textsc{ B.~Aharmim et al.},
  arXiv:1109.0763 .

     \bibitem{Collaboration:2011nga}%
  \textsc{G.~Bellini  et al.}, 
  Phys.\ Rev.\ Lett.  \textbf{108}, 051302 (2012).   
    

 
   \bibitem{Dziewonski:1981xy}%
  \textsc{A.~M.~Dziewonski}, and
  \textsc{D.~L.~Anderson},  
  Phys.\ Earth Planet.\ Interiors  \textbf{25}, 297 (1981).
  

   \bibitem{Ermilova:1986ph}%
  \textsc{V.~K.~Ermilova}, 
  \textsc{V.~A.~Tsarev}, and
  \textsc{V.~A.~Chechin}, 
  JETP Lett. \textbf{43}, 453 (1986)
  [Pisma Zh.\ Eksp.\ Teor.\ Fiz. \textbf{43}, 353 (1986)].
  

\bibitem{akres}%
\textsc{E. Kh. Akhmedov}, 
Yad. Fiz.  \textbf{47}, (1988) 475 
[Sov. J. Nucl. Phys.  \textbf{47} (1988) 301]. 
 

\bibitem{Akhmedov:1998ui}
\textsc{E. Kh. Akhmedov}, 
  Nucl.\ Phys.\ B \textbf{538}, 25 (1999).


   \bibitem{Petcov:1998su}%
  \textsc{S.~T.~Petcov}, 
   Phys.\ Lett.\ B \textbf{434}, 321 (1998).
  

   \bibitem{Akhmedov:1998xq}%
  \textsc{E.~K.~Akhmedov et al.}, 
  Nucl.\ Phys.\ B  \textbf{542}, 3 (1999).
   
   
     \bibitem{Akhmedov:2008qt}%
  \textsc{E.~K.~Akhmedov}, 
   \textsc{M.~Maltoni}, and
   \textsc{A.~Y.~Smirnov},
  JHEP \textbf{0806}, 072 (2008).


 \bibitem{Dighe:1999bi}%
  \textsc{A.~S.~Dighe}, and 
  \textsc{A.~Y.~Smirnov},
Phys.\ Rev.\ D \textbf{62}, 033007 (2000).
    

    \bibitem{Schirato:2002tg}%
  \textsc{R.~C.~Schirato}, and 
  \textsc{G.~M.~Fuller},
 astro-ph/0205390 .


 \bibitem{Tomas:2004gr}%
  \textsc{R.~Tomas et al.}, 
  JCAP \textbf{0409}, 015 (2004).

  \bibitem{Dasgupta:2005wn}%
  \textsc{B.~Dasgupta}, and 
  \textsc{A.~Dighe},
Phys.\ Rev.\ D \textbf{75}, 093002 (2007).



  
  

 
 
 \bibitem{Loreti:1995ae} 
  \textsc{ F.~N.~Loreti et al.}, 
  Phys.\ Rev.\ D \textbf{ 52}, 6664 (1995).
 
 
 \bibitem{Fogli:2006xy} 
   \textsc{ G.~L.~Fogli}, 
    \textsc{ E.~Lisi}, 
     \textsc{ A.~Mirizzi}, and 
      \textsc{ D.~Montanino},
JCAP \textbf{ 0606}, 012 (2006).
 
\bibitem{Friedland:2006ta} 
     \textsc{A.~Friedland}, and 
    \textsc{ A.~Gruzinov},
  astro-ph/0607244.
 
 \bibitem{Kneller:2010sc} 
  \textsc{J.~P.~Kneller}, and 
  \textsc{C.~Volpe},
Phys.\ Rev.\ D \textbf{ 82}, 123004 (2010).
 
\bibitem{Lund:2013uta} 
  \textsc{ T.~Lund} and 
   \textsc{J.~P.~Kneller},
  arXiv:1304.6372.
 
  \bibitem{Duan:2009cd}%
  \textsc{H.~Duan}, and 
  \textsc{J.~P. Kneller},
J.\ Phys.\ G \textbf{36}, , 113201 (2009).

   
  \bibitem{Pantaleone:1992eq}%
  \textsc{ J.~T.~Pantaleone},
  Phys.\ Lett.\  B \textbf{287}, 128 (1992).
    

 \bibitem{Samuel:1993uw}%
  \textsc{  S.~Samuel},
  Phys. Rev. D \textbf{48}, 1462 (1993).

 \bibitem{Balantekin:2004ug}%
  \textsc{A.~B.~Balantekin}, and 
  \textsc{H.~Yuksel},
New J.\ Phys. \textbf{7}, 51 (2005).


 \bibitem{Duan:2006an}%
  \textsc{H.~Duan}, 
  \textsc{G.~M.~Fuller},
    \textsc{J. Carlson}, 
      \textsc{Y.-Z.~Qian}, 
Phys.\ Rev.\ D \textbf{74}, 105014 (2006).

 \bibitem{Duan:2010bg}%
  \textsc{H.~Duan}, and 
  \textsc{G.~M.~Fuller}, and
    \textsc{Y.~-Z.~Qian},
Ann.\ Rev.\ Nucl.\ Part.\ Sci. \textbf{60}, 569 (2010).



 \bibitem{Duan:2007mv}%
  \textsc{ H.~Duan},  
  \textsc{G.~M.~Fuller},
 \textsc{J.~Carlson}, and
  \textsc{Y.~-Z.~Qian},  
Phys.\ Rev.\ D \textbf{75}, .125005 (2007).

 \bibitem{Hannestad:2006nj}%
  \textsc{S.~Hannestad}, 
  \textsc{G.~G.~Raffelt},
   \textsc{G.~Sigl }, and 
  \textsc{Y.~Y.~Y.~Wong}, 
Phys.\ Rev.\  D \textbf{ 74}, 105010 (2006)
  [Erratum-ibid.\  Phys.\ Rev.\  D \textbf{76}, 029901 (2007)].

\bibitem{Raffelt:2007cb}%
  \textsc{G.~G.~Raffelt}, 
  and \textsc{A.~Y.~Smirnov}, 
Phys.\ Rev.\ D \textbf{76}, 081301 (2007)
  [Erratum-ibid. Phys.\ Rev.\ D  \textbf{77}, 029903 (2008)].

 \bibitem{Galais:2011gh}%
  \textsc{S.~Galais}, and 
  \textsc{C.~Volpe},
Phys.\ Rev.\ D \textbf{84}, 085005 (2011).



 \bibitem{EstebanPretel:2008ni}%
  \textsc{A.~Esteban-Pretel et al.},
Phys.\ Rev.\ D \textbf{78}, 085012 (2008).

 \bibitem{Chakraborty:2011nf}%
  \textsc{S.~Chakraborty et al.}, 
Phys.\ Rev.\ Lett. \textbf{107}, 151101 (2011).

 \bibitem{Janka:2012wk}%
  \textsc{H.~-T.~Janka}, 
Ann.\ Rev.\ Nucl.\ Part.\ Sci. \textbf{62}, 407 (2012).



 \bibitem{Cherry:2012zw}%
  \textsc{J.~F.~Cherry et al.}, 
Phys.\ Rev.\ Lett. \textbf{108}, 261104 (2012).
 

 \bibitem{Balantekin:2006tg}%
 \textsc{A.~B.~Balantekin}, and
  \textsc{Y.~Pehlivan},
J.\ Phys.\ G \textbf{34},  47 (2007).
 
 
 \bibitem{Volpe:2013uxl}%
  \textsc{C.~Volpe}, 
    \textsc{D.~V\"a\"an\"anen}, and
      \textsc{C.~Espinoza,}
  Phys. Rev. D \textbf{87},  113010 (2013),
  arXiv:1302.2374. 
 

 \bibitem{Surman:2005kf}%
  \textsc{R.~Surman},  
  \textsc{G.~C.~McLaughlin},  and
    \textsc{W.~R.~Hix}, 
Astrophys.\ J. \textbf{643}, 1057 (2006).


 \bibitem{Kizivat:2010ea}%
  \textsc{L.~-T.~Kizivat et al.}, 
Phys.\ Rev.\ C \textbf{81}, 25802 (2010).

  \bibitem{Malkus:2012ts}%
  \textsc{A.~Malkus et al.}, 
Phys.\ Rev.\ D \textbf{86}, 085015 (2012).



 \bibitem{Dolgov:2002wy}%
  \textsc{A.~D.~Dolgov}, 
Phys.\ Rept. \textbf{370},  333 (2002).


 \bibitem{Iocco:2008va}%
  \textsc{F.~Iocco et al.}, 
Phys.\ Rept. \textbf{472}, 1 (2009).


 \bibitem{Hamann:2010bk}%
  \textsc{J.~Hamann et al.}, 
Phys.\ Rev.\ Lett. \textbf{105}, 181301 (2010).

\bibitem{Ade:2013ktc} 
  \textsc{ P.~A.~R.~Ade  et al.} , 
  arXiv:1303.5062.

\bibitem{Weinberg:1962zza}
 \textsc{S.~Weinberg}
  Phys.\ Rev.  \textbf{ 128}, 1457 (1962).



 \bibitem{Cocco:2007za}%
  \textsc{A.~G.~Cocco}, 
  \textsc{G.~Mangano},  and
  \textsc{M.~Messina}, 
  JCAP  \textbf{0706}, 015 (2007).


\bibitem{Lazauskas:2007da} 
 \textsc{ R.~Lazauskas}, 
 \textsc{P.~Vogel}, and 
 \textsc{C.~Volpe},
   J.\ Phys.\ G \textbf{ 35}, 025001 (2008).

 \bibitem{Davidson:2008bu}%
  \textsc{S.~Davidson et al.}, 
Phys.\ Rept. \textbf{466}, 105 (2008).

\bibitem{Minakata:1999ze}
 \textsc{H.~Minakata}, and 
  \textsc{S.~Watanabe}, 
Phys.\ Lett.\ B \textbf{468}, 256 (1999).

\bibitem{Akhmedov:2002zj} 
\textsc{E.~K.~Akhmedov}, 
\textsc{C.~Lunardini}, and
\textsc{A.~Y.~Smirnov}, 
Nucl.\ Phys.\ B  \textbf{643}, 339 (2002).

\bibitem{Balantekin:2007es}%
  \textsc{A.~B.~Balantekin}, 
   \textsc{J.~Gava}, and
    \textsc{C.~Volpe}, 
Phys.\ Lett.\ B \textbf{662}, 396 (2008).

\bibitem{Kneller:2009vd} 
  \textsc{J.~P.~Kneller}, and
  \textsc{G.~C.~McLaughlin},
Phys.\ Rev.\ D \textbf{80}, 053002 (2009).


\bibitem{Gava:2008rp} 
 \textsc{J.~Gava}, and
  \textsc{C.~Volpe}, 
Phys.\ Rev.\ D \textbf{78}, 083007 (2008).



\bibitem{deGouvea:2013zp} 
  \textsc{A.~de Gouvea},
   and \textsc{S.~Shalgar},
   JCAP \textbf{ 1304}, 018 (2013)].

\bibitem{Gava:2010kz} 
  \textsc{J.~Gava}, and
    \textsc{C.~Volpe},
Nucl.\ Phys.\ B  \textbf{837}, 50 (2010).

 \bibitem{Dighe:2003be}%
  \textsc{A.~S.~Dighe}, 
   \textsc{M.~T.~Keil},  and
    \textsc{G.~G.~Raffelt}, 
JCAP \textbf{0306}, 005 (2003).

\bibitem{Borriello:2012zc} 
    \textsc{E.~Borriello et al.}
   Phys.\ Rev.\ D \textbf{  86}, 083004 (2012).


 \bibitem{Kachelriess:2004ds} 
   \textsc{JM.~Kachelriess et al.}, 
  Phys.\ Rev.\ D   \textbf{ 71}, 063003 (2005).




\bibitem{Serpico:2011ir} 
  \textsc{P.~D.~Serpico et al.},
 Phys.\ Rev.\ D \textbf{85}, 085031 (2012)


\bibitem{Barger:2005it}%
  \textsc{V.~Barger}, 
 \textsc{P.~Huber}, and
  \textsc{D.~Marfatia}, 
Phys.\ Lett.\ B \textbf{ 617}, 167 (2005).

\bibitem{Gava:2009pj}
  \textsc{J.~Gava}, 
    \textsc{J.~Kneller}, 
  \textsc{C.~Volpe}, and 
  \textsc{G.~C.~McLaughlin}
Phys.\ Rev.\ Lett. \textbf{103}, 071101 (2009).



 \bibitem{snowglobes}%
 http://www.phy.duke.edu/~schol/snowglobes/

 \bibitem{Scholberg:2012id}%
  \textsc{K.~Scholberg}, 
  Ann.\ Rev.\ Nucl.\ Part.\ Sci. \textbf{62}, 81 (2012).


\bibitem{Volpe:2003fi} 
    \textsc{C.~Volpe},
J.\ Phys.\ G  \textbf{30}, L1 (2004).

 \bibitem{Avignone:2003ep}%
  \textsc{F.~T.~Avignone}, and 
  \textsc{Y.~.V.~Efremenko},
J.\ Phys.\ G  \textbf{29}, 2615 (2003).

\bibitem{Volpe:2005iy} 
  \textsc{ C.~Volpe},
J.\ Phys.\ G \textbf{ 31}, 903 (2005).


\bibitem{Mezzacappa:2005ju}
 A.~Mezzacappa,
 Ann.\ Rev.\ Nucl.\ Part.\ Sci. {\bf 55}, 467 (2005).


\bibitem{Kotake:2005zn}
 K.~Kotake, K.~Sato and K.~Takahashi,
 Rept.\ Prog.\ Phys. {\bf 69}, 971 (2006).


\bibitem{Lund:2012vm} 
  \textsc{T.~Lund et al.}, 
  Phys.\ Rev.\ D \textbf{86}, 105031 (2012).



\bibitem{McLaughlin:1999pd} 
  \textsc{  G.~C.~McLaughlin}, 
    \textsc{J.~M.~Fetter}, 
      \textsc{A.~B.~Balantekin}, and 
        \textsc{G.~M.~Fuller},
  Phys.\ Rev.\ C \textbf{ 59}, 2873 (1999).

\bibitem{Dasgupta:2011jf} 
  B.~Dasgupta, E.~P.~O'Connor and C.~D.~Ott,
  Phys.\ Rev. D {\bf 85}, 065008 (2012)
  [arXiv:1106.1167 [astro-ph.SR]].

 \bibitem{Collaboration:2011nsa}
  \textsc{M.~Ageron et al.}, 
Nucl.\ Instrum.\ Meth.\ A \textbf{656}, 11 (2011).

\bibitem{Kappes:2012hk}%
  \textsc{A.~Kappes}, 
\ Phys.\ Conf.\ Ser. \textbf{09}, 012014 (2013).

 \bibitem{HE-icecube}
 see the talk by N. WHITEHORN (Icecube collaboration) at "The IceCube Particle Astrophysics Symposium", May 13th-15th, Wisconsin-Madison.




 \bibitem{AdrianMartinez:2012ph}%
  \textsc{ S.~Adrian-Martinez et al.}, 
Phys.\ Lett.\ B \textbf{714}, 224 (2012).


 \bibitem{Gross:2013iq}%
  \textsc{A.~Gross}, 
  arXiv:1301.4339 [hep-ex].
 
 
 \bibitem{FernandezMartinez:2010am} 
    \textsc{E.~Fernandez-Martinez}, 
      \textsc{G.~Giordano}, 
        \textsc{O.~Mena}, and 
          \textsc{I.~Mocioiu},
  Phys.\ Rev.\ D \textbf{ 82}, 093011 (2010).
  
 
 \bibitem{Mena:2008rh} 
    \textsc{O.~Mena}, 
      \textsc{I.~Mocioiu}, and 
        \textsc{S.~Razzaque},
   Phys.\ Rev.\ D \textbf{ 78}, 093003 (2008).
 
 
  \bibitem{Akhmedov:2012ah}%
  \textsc{E.~K.~Akhmedov}, 
  \textsc{S.~Razzaque}, and
    \textsc{A.~Y.~Smirnov}, 
JHEP  \textbf{ 02}, 082 (2013)
  [JHEP  \textbf{ 1302}, 082 (2013)]
 

 
\bibitem{Serpico:2005sz} 
 \textsc{  P.~D.~Serpico}, and 
  \textsc{M.~Kachelriess},
   \textsc{ Phys.\ Rev.\ Lett.} \textbf{ 94}, 211102 (2005).


\bibitem{Winter:2006ce} 
 \textsc{ W.~Winter},
    Phys.\ Rev.\ D \textbf{74}, 033015 (2006).


 \bibitem{Meloni:2012nk}%
  \textsc{D.~Meloni}, and
  \textsc{ T.~Ohlsson },
  Phys.\ Rev.\ D \textbf{86}, 067701 (2012).


 
 \bibitem{Esmaili:2012nz}%
  \textsc{A.~Esmaili et al.}, 
JCAP \textbf{1211}, 041 (2012).
 
 \bibitem{Razzaque:2012tp}%
  \textsc{S.~Razzaque}, and
  \textsc{A.~Y.~Smirnov}, 
Phys.\ Rev.\ D \textbf{85}, 093010 (2012).

 
 \bibitem{Abbasi:2010kx}%
  \textsc{R.~Abbasi et al.}, 
  Phys.\ Rev.\ D \textbf{ 82}, 112003 (2010).



\bibitem{Longo:1987ub} 
    \textsc{M.~J.~Longo},
  Phys.\ Rev.\ D {\bf 36}, 3276 (1987).

 \bibitem{Katori:2012hc}%
 \textsc{  T.~Katori},
  arXiv:1211.7129.
  
 \bibitem{Diaz:2011tx} 
    \textsc{J.~S.~Diaz},
  arXiv:1109.4620.




\end{thebibliography}
\end{document}